# LARES SUCCESFULLY LAUNCHED IN ORBIT: SATELLITE AND MISSION DESCRIPTION

Antonio Paolozzi[1]     Ignazio Ciufolini[2]

[1]Scuola di Ingegneria Aerospaziale, and DIAEE, Sapienza Università di Roma
[2]Dip.Ingegneria dell'Innovazione, Università del Salento, Lecce and Centro Fermi, Roma

**Abstract**

On February 13$^{th}$ 2012, the LARES satellite of the Italian Space Agency (ASI) was launched into orbit with the qualification flight of the new VEGA launcher of the European Space Agency (ESA). The payload was released very accurately in the nominal orbit. The name LARES means LAser RElativity Satellite and summarises the objective of the mission and some characteristics of the satellite. It is, in fact, a mission designed to test Einstein's General Relativity Theory (specifically 'frame dragging' and Lense-Thirring effect). The satellite is passive and covered with optical retroreflectors that send back laser pulses to the emitting ground station. This allows accurate positioning of the satellite, which is important for measuring the very small deviations from Galilei-Newton's laws. In 2008, ASI selected the prime industrial contractor for the LARES system with a heavy involvement of the universities in all phases of the programme, from the design to the construction and testing of the satellite and separation system. The data exploitation phase started immediately after the launch under a new contract between ASI and those universities. Tracking of the satellite is provided by the International Laser Ranging Service. Due to its particular design, LARES is the orbiting object with the highest known mean density in the solar system. In this paper, it is shown that this peculiarity makes it the best proof particle ever manufactured. Design aspects, mission objectives and preliminary data analysis will be also presented.

**Keywords:** satellite design and manufacturing; laser ranged satellite; scientific space mission; tests of general relativity; orbital determination; frame-dragging.

## 1. Introduction

Satellite Laser Ranging (SLR) is the most accurate technique available today to measure distances to artificial satellites and is therefore suitable to detect deviations from classical Galilei-Newton mechanics due to General Relativity in the trajectories of orbiting bodies. The idea of using laser ranged data for fundamental physics dates back to the 1960's and was applied by the Apollo 11 mission with the deployment of a laser ranging array on the Moon's surface by the astronaut Neil Armstrong [1]. However, the first application of laser ranging was in the field of Earth sciences. In fact, six laser ranged satellites were launched before 1969, starting with Beacon-B (NASA 1964) [2]. Later, in 1976, a milestone in SLR was set by NASA with the launch of LAGEOS (LAser GEodynamic Satellite) and in 1992 by NASA and ASI with the launch of LAGEOS 2. In 1984 [3, 4], it was proposed to test the frame dragging and Lense-Thirring effect using two LAGEOS-type satellites. Frame dragging was predicted by Einstein in

1913 and the Lense-Thirring effect was derived from General Relativity by the two Austrian physicists Josef Lense and Hans Thirring [5]. In 1989 [6,7] and in 1996 [8], it was proposed to use LAGEOS 1 and LAGEOS 2 to perform that accurate measurement. In particular in 1989 [6, 7], it was proposed to change the decided inclination of LAGEOS 2 to 70 degrees so that its orbit would be supplementary to that of LAGEOS, with both satellites orbiting at about 6000 km altitude. With this configuration, the error in the measurement of frame dragging due to all the even zonal harmonics would have been eliminated. Unfortunately the final orbital inclination chosen for LAGEOS 2 was not optimal for General Relativity and the measurement was limited by the uncertainties in the even zonal harmonics $J_{2n}$ of the Earth's gravitational field of degree strictly higher than two [9-12]. For this type of satellite, a parameter important both for geodesy and fundamental physics is the cross sectional surface-to-mass ratio. The smaller this parameter, the closer is the behaviour of the satellite to an ideal test-particle. The LAGEOS satellites, in this respect, are considered the reference for scientists, as they had, until 2012, the smallest value ever reached by an artificial orbiting object: $7 \cdot 10^{-4}$ m$^2$/kg. With LARES, this value has improved by a factor of 2.7 and it is, in fact, the best test-particle ever manufactured [13, 14].

In four years, the satellite [15-17], the separation system [18-22] and the Cube Corner Reflectors (CCRs) [23, 24] were designed, manufactured [25-29], and tested. Optics related to CCRs is quite unusual, and it was particularly difficult to prepare a laboratory for testing their properties under space environment [30, 31]. However our group acquired some experience with a space CCR during the Specular Point-like Quick Reference (SPQR) experiment conducted on the International Space Station [32, 33]. SPQR proved that from ground based photographs, damages on spacecraft could be detected if the atmospheric disturbance is evaluated (by a reflected laser beam from a CCR on board the spacecraft) and removed with proper image processing. With such a background it was possible to deal with those issues in the limited timeframe available. One of the final decisions to be taken was relevant to the metallic thermo-optical surface properties of the satellite. In particular it had to be decided whether to paint the satellite or to perform a surface treatment [25, 34]. The tests proved [30] that only a cleaning of the satellite was necessary.

The basic idea of the LARES experiment is to accurately reconstruct the actual orbit from laser ranging data and to compare it with the theoretical one obtained using all the forces and perturbations acting on the satellite [35, 36]. Very important in this comparison is the level of accuracy reached in the knowledge of the classical forces and perturbations [12, 37–43]. In 1998 a first version of the LARES mission was presented in response to a call issued by ASI in 1997. Already at that time the importance of the mission was recognized, as it was among the six proposals selected for a phase A study [44]. In 1999, for the development of the next phases B to E, only one of the six proposals selected was approved for launch: AGILE (Astrorivelatore Gamma ad Immagini LEggero). It is a gamma and X-ray telescope of ASI, launched with the Indian launcher PSLV-C8 on April 23, 2007. The group continued to work on the idea of a new laser ranged satellite for several years [45-52]. Finally, on the 7[th] of February 2008, ASI awarded the contract to CGS (Compagnia Generale dello Spazio, formerly Carlo Gavazzi Space) for the B-C-D phases. The main subcontractors were the Universities of Sapienza and Salento, which were in charge of the design of the satellite, separation system, pieces of Mechanical and Ground Support Equipment (MGSE), thermal and optical tests, the pre-load measurement on the separation system and the scientific supervision. Other industrial subcontractors were SAB Aerospace (Società Aerospaziale Benevento) and Telematics Solutions (formerly part of Rheinmetall Italia) for the structural components and TEMIS (Transport Electronic Mechatronic

Integrated Systems) for the avionics components. INFN (Istituto Nazionale di Fisica Nucleare) provided support for the selection of the CCRs, which were acquired from Zeiss in Germany.

## 2. Frame Dragging

In General Relativity test-gyroscopes define the axes of the local inertial frames [53] and they are dragged by mass-energy currents, hence the name "frame-dragging" given to this phenomenon. Whereas in electromagnetism an electric current generates a magnetic field, there is no such analogue in Newtonian gravitational theory. Nevertheless, in the gravitational theory of General Relativity, a current of mass-energy generates a gravitomagnetic field [1, 53-55]. In the limit of weak-field and slow motion, the governing equations are formally identical with those of electromagnetism with the exception of a negative constant value (gravitation is attractive). However, this analogy is only formal and for Solar System phenomena the two fields are dramatically different in magnitude and extremely weak in the case of General Relativity, which makes this effect very difficult to detect around the Earth. Incidentally, this diverse magnitude explains why magnetic forces were discovered many centuries before gravitomagnetism.

Frame dragging has an effect on gyroscopes, test-particles, electromagnetic waves and clocks. For example, one clock co-rotating with respect to a central massive rotating body and another clock initially synchronized with it but counter-rotating with respect to the central rotating body will display a time difference upon encounter, the counter-rotating clock being retarded relative to the rotating one. This time difference is extremely small for two clocks counter-rotating around Earth. The situation is different when the gravitational field is very strong, such as around rotating black holes or supermassive black holes. In such cases, frame dragging is much stronger and the space-time deformation is so high that frame-dragging effects can be very large. Frame-dragging produces also a time-delay in the propagation time of electromagnetic waves counter-propagating with respect to the rotation of a central body [56,57]. The effect of frame-dragging on a gyroscope was derived by Pugh and Schiff [58, 59] Finally, frame-dragging produces the precession of the orbit of a test-particle. Both the orbital angular momentum vector and the Runge-Lenz vector, determining the pericenter of the orbiting test-particle, have a precession. In the case of weak gravitational field and slow motion, the Lense-Thirring effect [5] describes the frame-dragging precession of the node and pericenter of a test-particle (see the next section).

## 3. The Science Objective

Although LARES is also used by space geodesists as an Earth science satellite, its main objective is to contribute to fundamental physics. In particular, its orbit has been optimized to reduce the errors in the measurement of frame dragging due to the Earth's even zonal harmonics, i.e. those deviations of the Earth's gravity field from spherical symmetry that are symmetric with respect to the Earth's figure axis and to its equatorial plane. Indeed, these spherical harmonics produce large secular effects on the LARES and LAGEOS nodal lines (intersection of Earth's equatorial and orbital planes). Error minimization has been achieved by choosing an inclination as close as possible to 70°. This inclination was also considered safe from the launch point of view and was only slightly changed to 69.5° when the mission parameters were finally set. The uncertainties associated with the even zonal harmonics in unit of Lense-Thirring effect are reported in Table 1. It is clear, for instance, that the uncertainty due to the harmonic of degree 2 and order 0, the

Earth's quadrupole moment, $J_2$, is for LAGEOS 1.6 times the Lense-Thirring effect. This means that frame dragging cannot be directly measured using only the LAGEOS satellite. By using instead the orbital data of LAGEOS and LAGEOS 2 it was possible to eliminate $J_2$, reaching an accuracy that was basically limited by the uncertainty of $J_4$ and higher degree even zonal harmonics plus the uncertainties due to the non-gravitational perturbations and to Earth's tides. The best value obtained so far is about 10% [9,12,39]. By adding the LARES orbital data it will be possible to eliminate also the effects of $J_4$, thus allowing the achievement of about 1% accuracy. It has to be considered that the accurate measurement will require a number of years during which the effects of some periodical perturbations will average to zero and the knowledge of the gravitational field of Earth will improve due to the analysis of new data from the GRACE (Gravity Recovery And Climate Experiment) and GOCE (Gravity field and steady-state Ocean Circulation Explorer) missions. We finally point out that the concerns raised by one author, see, e.g., [60] are not based on solid grounds as clearly addressed and answered in a number of papers, see, e.g., [12,40,41] and references therein. The comments of that author [60] in regard to the errors due to the higher degree even zonal harmonics and to other conceivable error sources in the LARES experiment are often even contradicting each other as pointed out, e.g., in [41].

## 4. LARES System

To interface the launch vehicle with the satellite, a LARES system has been designed, as shown in Fig. 1. The main components are the satellite, the separation system, the support system with a total mass of 598.2 kg, avionics and harness, including two video cameras with a total of mass of 41.97 kg. Additional elements are eight piggy back passengers with a total mass of about 45 kg. These were provided by universities from all over Europe.

*4.1 LARES satellite*
LARES is a spherical passive satellite designed with the lowest surface-to-mass ratio ever achieved. This design minimizes the effect of the non-gravitational perturbations acting on the surface of the satellite, thus allowing it to be considered the best artificial test-particle available today for fundamental physics and geodesy. The satellite radius is 182 mm, and 92 CCRs are mounted on its surface through a mounting system (Fig. 2) that has been chosen from among several options to reduce thermal thrust perturbations to a minimum. The satellite body is a massive single piece of tungsten alloy of density 18000 kg/m$^3$. The mean density of the satellite decreases to 15300 kg/m$^3$ due to the cavities, the CCRs and their plastic mounting rings. Many challenges were faced during manufacturing, mainly related to the tight tolerances required on a material with higher hardness than an aluminium alloy. The material, never before used for an aerospace structure, was obtained by the liquid phase sintering process [61] in which tungsten particles are embedded in a Ni-Cu matrix. This aspect was particularly critical for manufacturing the screw threads: it was necessary to increase the initial design diameter from 2 mm to 3 mm and to change the manufacturing technology from lathe machining to plastic deformation. Another very difficult task was to machine the hemispherical cavities, designed to interface with the separation system. More details will be given in section 4.2.

*4.2 Separation and support systems*
From a structural point of view the separation system and its interface with the satellite was the most demanding [19, 62]. The development of the LARES system paralleled the final part of the VEGA launcher construction, and initially the acceleration level requirements were too high and they were not compatible with the scientific requirements of the satellite. In particular to maintain LARES during launch, protruding parts were not acceptable on the satellite exterior, nor were significant modifications on the spherical surface. Fortunately the final design acceleration levels were much lower than initially decided: 5 g axial and 3.5 g lateral. But in spite of this reduction, the challenge of maintaining a 400 kg satellite with the abovementioned requirements was quite high. The solution selected was to machine a number of hemispherical cavities on the equatorial line of the satellite. A pin was engaged in each cavity to maintain the satellite in place. The number and size of the cavities needed to be as small as possible and were reduced by pushing tolerances to the technological limit. This is because the pressure at the contact point will increase as a function of the difference of the two radii of pin and cavity. For the pin, the tolerance was limited to 0.04 mm, while for the cavity (more difficult to manufacture), it was not possible to be tighter than 0.1 mm [20]. To give an idea of the dependence of local stress at the contact area as a function of the tolerances, we report in Fig. 3 what is obtained using the Hertz formula [63, pag 702, Tab 14.1 case 1c]. The formula provides the maximum stress and the area of the contact surface after deformation as a function of: (i) the elastic constants of the two materials, (ii) the radii of the two spherical surfaces in contact and (iii) the pushing force. The pressure at the contact area vanishes if the limit of zero difference in the two radii is reached and increases rapidly when the difference is only a few tenths of a millimetre. Specifically, the analytical values calculated using Hertz theory are 371.05 MPa for the nominal case (pin radius16.80 mm, cavity radius 17.00 mm) and 419.67 MPa for the worst geometric condition (maximum difference in pin and cavity radii within required tolerance, i.e. 0.24 mm) [21]. The input values used in the calculation are reported in Table 2. The number of hemispherical cavities has been decided based on the analytical results of Fig. 4, which reports the maximum contact pressure for the worst tolerance case. The blue thicker line refers to three cavities, the red thinner line to four cavities. The admissible value of stress is 481 MPa, i.e. 93% of the yield limit of the weaker material, as required by the European Cooperation for Space Standardization (ECSS) [64]. So in theory it would have been possible to use three cavities with a diameter of 33 mm because in this case we would have had a value just below the admissible stress. However, the FEM model of the pin/cavity interface provided a maximum value of the pressure [21] that was 8.5% higher than the analytical one, thus making this solution unacceptable. A better situation would have been obtained by increasing the diameter of the cavity to 37 mm. In this case, three cavities would have provided a maximum value of the contact stress of about 410 MPa, which from this point of view would have been acceptable. However, looking at the detail of the constructive drawing (Fig. 5), the CCR and hemispherical cavities are already very close to take the risk to increase the 34 mm diameter to 36mm. Thus the final decision was to use four pins although the mechanism with four actuators is less reliable than the solution with three. Indeed failure of two actuators will imply failure of separation and consequently of the whole science mission.

　　Two units of LARES satellite were manufactured: the flight unit and the Demonstration Model (DM), used for qualifying the separation system and the LARES system. The DM was manufactured using the same batch material as the flight unit and was identical to it but with only 20 CCR cavities completed. This brought its mass to about 400 kg, i.e. approximately 13 kg more than the flight unit. In case of any problem with the flight unit, the DM could have been easily machined according to the final constructive drawings to get a second flight unit. The four actuators are of the class of Non Explosive Actuators (NEA) and operate independently. This solution is different from the one used on LAGEOS 2, where a Marman clamp was cut by two redundant explosive pyrotechnical cutting devices, offering the advantage of working also in case of failure of one actuator. This solution could not be considered for LARES as the cost of a space-qualified Marman clamp was outside the allocated budget. Fig. 6 shows the satellite mounted on the separation system. The ejection system at the bottom consists of a single spring pushing with about 3700 N. This value is lower than the weight of the satellite (in the event of accidental actuation while on the ground, the satellite would not be ejected). Also, the ejection speed is about 0.75 m/s, which is necessary to meet some collision avoidance at the separation from the upper-stage.

The support system (Fig. 1) has been designed to maintain the satellite's centre of mass in a position closer to that of a typical passenger, that would be instead anchored directly to the last VEGA stage interface. As a result of this approach a complication arose due to the difficulty of maintaining the natural frequencies above the required values. The presence of bolted joints in the support system made the prediction of its modal characteristics difficult, so that a high value of safety factor was considered. This explains the high values obtained experimentally (Table 3).

*4.3 Avionics and Harness*

Besides providing the signals and power for LARES and the secondary payload separations, the avionics comprise several pieces of equipment for telemetry, environmental sensors and conditioners. Furthermore, an internal and an external camera have been added to the system.

## 5. Ground segment

The ground segment of the mission is provided by the International Laser Ranging Service (ILRS) [66], which coordinates 50 laser ranging stations distributed all over the world (Fig. 7) though only about 40 are really contributing data. One of these is the Matera Laser Ranging Observatory (MLRO), owned by ASI, which is one of the best stations for space geodesy in the world. Currently, ILRS tracks 39 targets: geodetic spherical satellites (10), navigation satellites (16), remote sensing satellites and scientific satellites. All of them carry CCRs with the exception of BLITS (Ball Lens In The Space), which is a single spherical experimental optical laser reflector. The measurement principle is based on the accurate measurement of the return time of flight of laser pulses. Also, the knowledge of the speed of light in the atmospheric portion involves the use of atmospheric models that evolved from the Marini-Murray model [67] to the Mendes-Pavlis model [68-70]. In order to track the satellite, the LARES scientific team is in charge of providing orbital predictions to the ILRS based on initial conditions that were given by the launch vehicle soon after the separation of LARES and later by the radar data from NORAD (**NOR**th American **A**erospace **D**efense Command). Currently, the process uses information from ILRS laser ranging data. The various tasks of laser ranged satellites imply different orbital parameters so that their altitudes range from 600 km to 24000 km. There are five more targets on the Moon, but only a few stations are capable of operating at that distance. The 10 geodetic satellites are spherical and passive and the only communication is based on the laser pulses reflected by the retroreflectors. This information, in spite of its simplicity, will be processed to reconstruct with centimetre accuracy the satellite orbit, thus allowing for the Lense-Thirring effect measurement. On the other hand, geodesists will exploit laser ranging data to monitor the position of the Earth's centre of mass and orientation of its rotational axis. In addition, the effect of global climate change, such as global ice melting, and tectonic movements can be investigated, and accurate determination of the terrestrial reference frame can be performed. LARES is currently tracked by 37 stations and several millions of observations are already available.

Laser ranging data are stored in full rate data according to two different formats: MERIT II (since the 1970's) and in the new Consolidated laser Ranging Data (CRD) since April 2008. The CRD format includes the normal points and has actually been used for this purpose since 2006. Normal points are reduced data from the full rate data. In Table 4, the stations that use the new format with the number of passes and full rate observations are reported. In Table 5, the time of last observation and the normal points are reported.

### 6. Preliminary data analysis

On the 13$^{th}$ of February 2012 at 7:00 am local time, LARES was launched successfully from the ESA spaceport in Kourou, French Guyana. After 55 minutes, LARES was separated. On Friday the 17$^{th}$ the first laser return was recorded. Since then, a very high number of data have become available (Tables 4 and 5). From the preliminary orbital analyses of the normal points we have observed that LARES approximates the behaviour of a test-particle better than the LAGEOS satellites, which were, until now, the best test-particles ever manufactured. That was somehow expected because of the particular design of LARES, which was specifically conceived to minimise the effects of the non-gravitational perturbations, though there were some concerns about the much lower orbit (1450 km vs. 5900 km). The single piece design and the lower CCR surface-to-metallic surface ratio by a factor of 1.65 reduce the thermal thrust perturbation. In particular, the surface-to-mass ratio, improved by a factor of about 2.7 with respect to LAGEOS, reduces the effects of all surface non-gravitational perturbations, such as atmospheric drag. A quantitative, though preliminary, analysis is reported in [72] and summarised below.

Although out-of-plane accelerations are the relevant ones for the measurement of frame dragging, the along-track accelerations provide a quality index for a test-particle. After modelling the known orbital perturbations, LAGEOS data analysis for 105 days (starting from 15$^{th}$ of February 2012) provided residual along-track acceleration in the range 1 to 2 x 10$^{-12}$ m/s$^2$, whereas LARES provided a value of less than 0.4 x 10$^{-12}$ m/s$^2$. As a result, the deviation from space-time geodesic motion is smaller for LARES, as reported in the semi-logarithmic diagram of Fig. 8. In Ref. [73] the time evolution of the orbital parameters is derived from the laser ranging observations; in Table 6 are reported the actual mean values of those parameters.

### 7. Conclusions

The first orbital analyses performed on laser ranging data of LARES prove that the innovative design adopted for the satellite was successful. The satellite performs as the best test-particle available today for fundamental physics studies. Also, the geodesy and geodynamics communities benefit from the use of this new laser ranging satellite. It is expected that in a number of years, thanks to the averaging out of the effects of some periodic perturbations, the accuracy of the measurement of frame dragging will improve by about one order of magnitude, providing an important benchmark in physics.

**Acknowledgments**

The authors thank the Italian Space Agency and in particular its President, Enrico Saggese, its former President Giovanni Bignami and its Technical Director, Mario Cosmo. The work is performed under the ASI contract n. I/034/12/0. The authors acknowledge the European Space Agency for providing the qualification launch and the International Laser Ranging Service for tracking and data distribution of the LARES satellite.

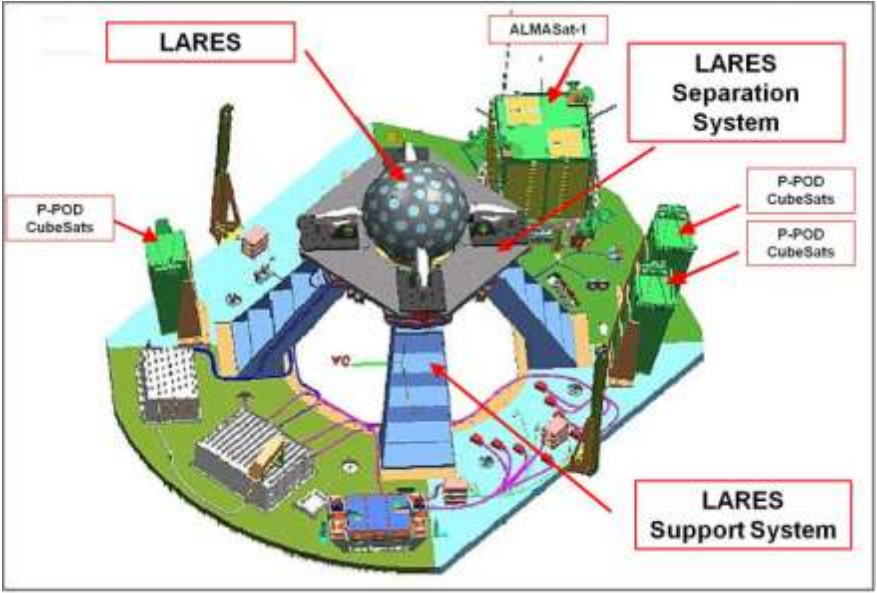

**Fig. 1.** LARES system.

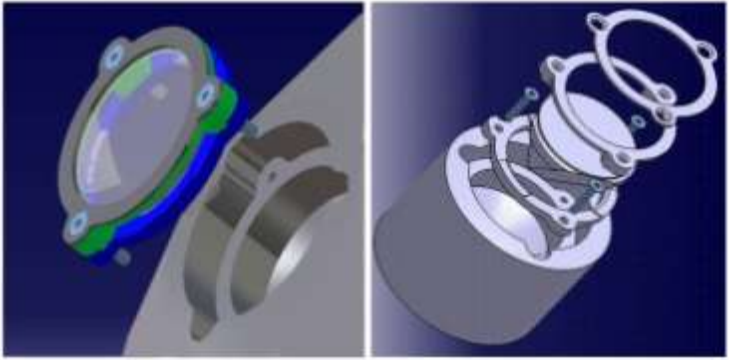

**Fig. 2.** CCR mounting system: assembled (left) and exploded (right) views.

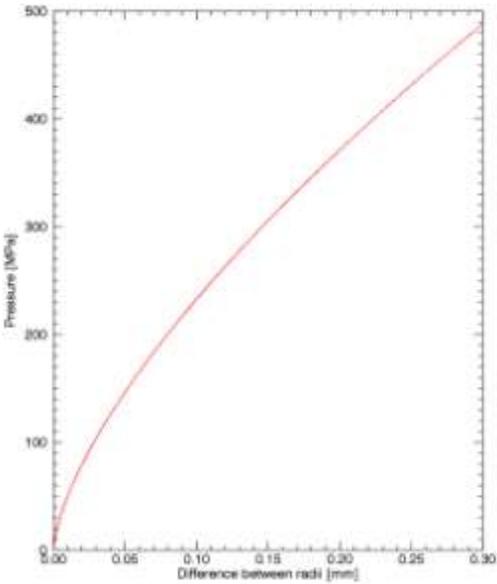

**Fig. 3.** Maximum stress at contact area at the satellite/separation system interface.

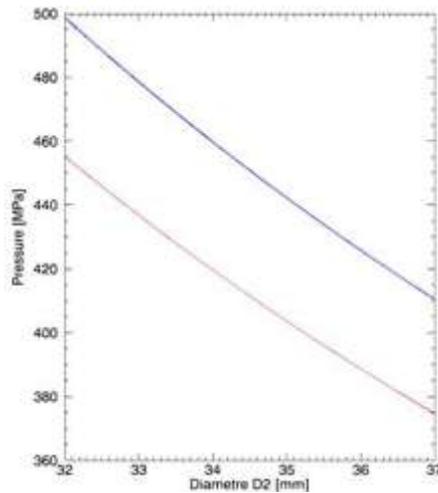

**Fig. 4.** Contact pressure as a function of hemispherical cavity diameter; four cavities (red thin line) three cavities (blue thick line). Worst tolerance case.

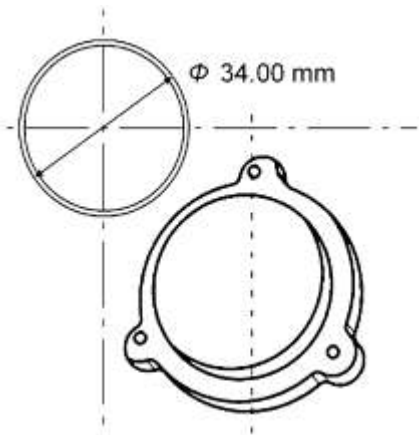

**Fig. 5.** Detail of LARES constructive drawing: hemispherical cavity (top left) and CCR cavity (bottom right).

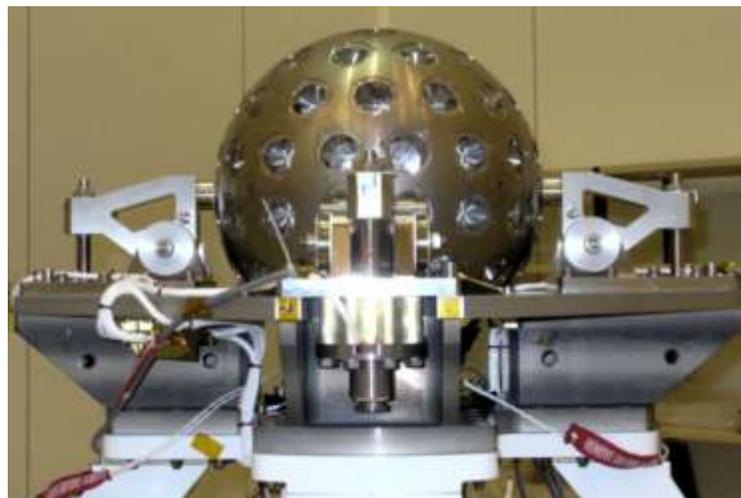

**Fig. 6.** LARES Flight Unit mounted on separation system.

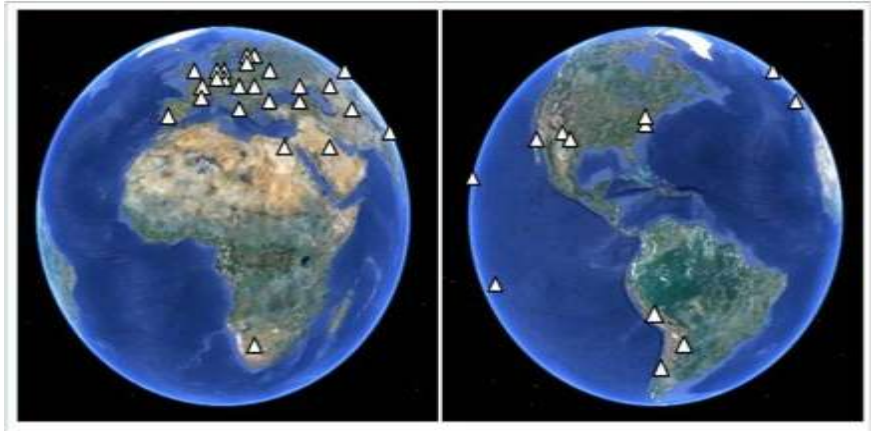

**Fig. 7.** Location of the ILRS laser ranging stations.

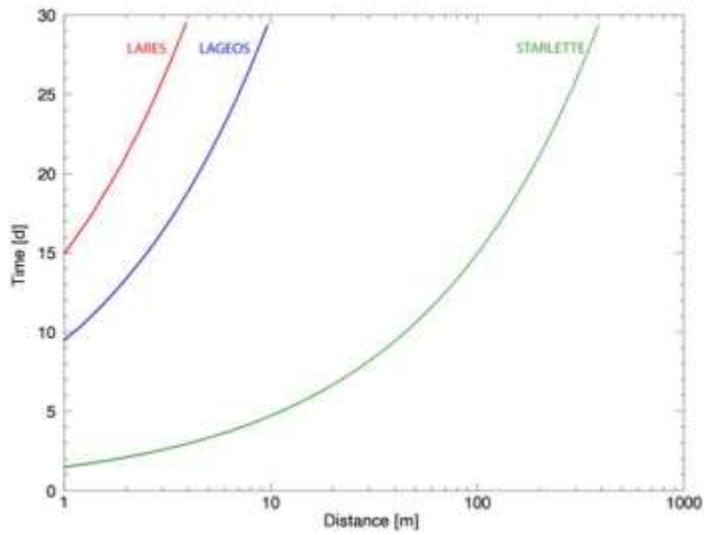

**Fig. 8.** Along-track deviations from geodesic motion of some laser ranged satellites after modelling the known orbital perturbations.

**Table 1.**
Uncertainties on the node (EIGEN-GRACE02S Earth gravitational field), in units of the frame-dragging effect on the node.

| Even zonal | LAGEOS | LAGEOS 2 | LARES |
|---|---|---|---|
| 2 0 | 1.6 | 2.9 | 2.1 |
| 4 0 | 0.059 | 0.021 | 0.18 |

**Table 2.**
Material properties of satellite and separation system (E=Young modulus, ν=Poisson ratio, r=radius).

| **Steel** | | **Tungsten** | |
|---|---|---|---|
| E (GPa) | 210 | E (GPa) | 345 |
| ν | 0.27 | ν | 0.31 |
| Pin, $r_{min}$ (mm) | 16.76 | Cavity, $r_{min}$ (mm) | 16.90 |
| Pin, $r_{max}$ (mm) | 16.80 | Cavity, $r_{max}$ (mm) | 17.00 |

**Table 3.**
LARES Support System natural frequencies, Ref. [65].

| Frequency | Requirement (Hz) | Measured (Hz) |
|---|---|---|
| First Lateral | ≥ 60 | 144 |
| First Longitudinal | ≥ 15 | 48 |

**Table 4.**
Current ILRS tracking statistics of full rate data (CRD), Ref. [71].

| **Stations** | **Passes** | **Observations** |
|---|---|---|
| Golosiiv | 96 | 5434 |
| Simeiz | 139 | 15055 |
| Katzively | 206 | 32307 |
| Changchun | 531 | 6314799 |
| Concepcion (1) | 4 | 1267 |
| Concepcion (2) | 168 | 143260 |
| San Juan | 262 | 46804 |
| Zimmerwald | 251 | 738479 |
| Shanghai | 77 | 1313631 |
| San Fernando | 120 | 21448 |
| Mt Stromlo | 387 | 338700 |
| Herstmonceux | 333 | 545381 |
| Potsdam | 285 | 3415830 |
| Grasse | 111 | 76735 |
| Matera | 468 | 468527 |
| Wettzell | 319 | 230112 |

**Table 5.**
Current ILRS tracking statistics of normal points data (CRD), Ref. [71].

| Stations | Last observation | Normal Points |
|---|---|---|
| Golosiiv | 04/02/2013 19:48 | 780 |
| Simeiz | 04/01/2013 02:30 | 1321 |
| Katzively | 19/01/2013 17:07 | 2088 |
| Changchun | 15/02/2013 16:06 | 3628 |
| Concepcion(1) | 23/11/2012 07:19 | 13 |
| Concepcion(2) | 14/02/2013 08:05 | 725 |
| San Juan | 16/02/2013 06:00 | 3191 |
| Zimmerwald | N.A. | 8404 |
| Shanghai | 03/02/2013 17:10 | 524 |
| San Fernando | 15/02/2013 23:44 | 708 |
| Mt Stromlo | 14/02/2013 00:57 | 4190 |
| Herstmonceux | 14/02/2013 22:52 | 4295 |
| Potsdam | 12/02/2013 13:03 | 4891 |
| Grasse | 13/02/2013 23:59 | 1765 |
| Matera | 11/02/2013 22:05 | 5339 |
| Wettzell | 12/02/2013 23:00 | 4859 |

**Table 6.**
LARES orbital parameters.

| Orbital parameters | Nominal value | Actual value |
|---|---|---|
| Semi-major axis | 7825 km | 7820 km |
| Inclination | 69.5° | 69.5° |
| Eccentricity | 0 | 0.0007 |